\documentclass[twocolumn,showpacs,showkeys,preprintnumbers,amsmath,amssymb,superscriptaddress]{revtex4}
\usepackage{graphicx}% Include figure files
\usepackage{dcolumn}% Align table columns on decimal point
\usepackage{bm}% bold math

\begin{document}

\title{Phase Bifurcations of Strongly Correlated Electron Gas \\ at the Conditions of dHvA Effect}% Force line breaks with \\

\author{Nathan Logoboy}

\email{logoboy@phys.huji.ac.il}

\affiliation{Grenoble High Magnetic Field Laboratory, MPI-FKF and
CNRS P.O. 166X, F-38042 Grenoble Cedex 9, France}

\affiliation {The Racah Institute of Physics, The Hebrew University
of Jerusalem, 91904 Jerusalem, Israel}

\author{Walter Joss}
\affiliation{Grenoble High Magnetic Field Laboratory, MPI-FKF and
CNRS P.O. 166X, F-38042 Grenoble Cedex 9, France}

\affiliation {Universit$\acute{e}$ Joseph Fourier, B.P. 53, F-38041
Grenoble Cedex 9, France}

\date{\today}% It is always \today, today,

\begin{abstract}

In a framework of catastrophe theory we investigate the
equilibrium set for the system of strongly correlated electron gas
at the conditions of dHvA effect and show that the discontinuities
accompanied the diamagnetic phase transition (DPT) is handled by
Riemann-Hugoniot catastrophe. We show that applicability of the
standard condition for observation of DPT $a\ge$1 where $a$ is the
differential magnetic susceptibility is valid only in the nearest
vicinity of triple degenerate point corresponding to the center of
dHvA period, but for arbitrary value of magnetic field in every
period of dHvA oscillations this condition is modified in accordance
with the bifurcation set of cusp catastrophe. While at the center of dHvA period the
symmetric supercritical pitchfork bifurcation gives rise to the
second order phase transition on temperature, the deviation of
magnetic field from the value corresponding to the center
of dHvA period results in the change of the phase transition order
from the second to the first one both on temperature and magnetic
field. In the framework of developed theory we obtain good agreement
with available experimental data.

\end{abstract}



\maketitle

\section{\label{sec:Introduction}Introduction}

The instability of an electron gas due to strong correlations at the
conditions of dHvA effect resulting in diamagnetic phase transition
(DPT) into inhomogeneous diamagnetic phase (IDP) with formation of
Condon domains (CDs) \cite{Condon}, \cite{Condon_Walstedt} is
intensively studied both theoretically and experimentally
\cite{Kramer1}-\cite{Logoboy3}. The realization of intrinsic
structure of IDP is governed by the competition between long-range
dipole-dipole interaction and short-range interaction related to the
positive interface energy with typical magnetic length of Larmor
radius. The observed IDP in Ag by splitting of NMR line
\cite{Condon_Walstedt} was identified as periodic domain structure
with alternating in neighboring domains one-component magnetization.
The further experiments on observation of IDP by method of $\mu$SR
spectroscopy revealed the arise of diamagnetic instability in Be,
Sn, Al, Pb and In \cite{Solt1}-\cite{Solt3}, giving strong support
to the idea of formation of CD structure \cite{Condon}.

The experimental demonstration of existence of IDP in normal metals
\cite{Condon_Walstedt},\cite{Solt1}-\cite{Solt3} is very convincing
and support the basic ideas of the theory of DPT developed in
\cite{Condon}, \cite{Shoenberg}, but the full
understanding of the properties of IDP is still lacking. The attempts of the
quantitative analysis of the data on the basis of model of
plane-parallel band patterns reveal the contradictions \cite{Solt3},
put questions on correctness of applicability of demagnetizing
coefficient for adequate description of shape-dependent properties
of IDP \cite{Solt3}, \cite{Logoboy1}, \cite{Plummer} and raised
important unanswered questions concerning the type of the
diamagnetic ordering, irreversibility of DPT and morphology of the
domain patterns.

There are striking similarities between IDP in normal metals and
other strongly correlated systems which undergo phase transition on
temperature and magnetic field with formation of complex macroscopic
patterns, e. g. the type-I superconductors and thin magnetic films.
The different technics, including the powder pattern and
magneto-optic methods, successfully used for observation of
intermediate state of type-I superconductors revealed a very rich
structure \cite{Livingstone} which in spite of all it complexity,
amazingly reminds the variety of domain structures in thin film
\cite{Hubert}. Recent experimental observation of CDs by a set of
micro Hall probes at the surface of a plate-like sample of Ag
\cite{Kramer1} demonstrates the formation of complex structure which
differs from the expected regular CD patterns separated by
plane-parallel DWs \cite{Condon} and calls for further development
of the experimental techniques to study the IDP. The first
experimental observation of such an exotic phenomena as diamagnetic
hysteresis by complex method including direct measurements by Hall
probe, standard ac method with different modulation levels,
frequencies and magnetic field ramp rates \cite{Kramer2} allows to
reconstruct the magnetization reversal in Be and confirms the
possibility of the first order phase transition. The giant
parametric amplification of non-linear response which proves to be
sensitive method for investigation of phase transitions in
superconductors was observed in a single crystal of Be in quantizing
magnetic field \cite{Tsindlekht1}, where the measured output signal
amazingly follows the shape of the DPT diagram and shows the
asymmetry and shift to the upper edge of the dHvA period in
ascending magnetic field justifying the possibility of the first
order DPT and necessity of further development of the theory of DPT.

Motivated by recent data on observation of DPT instability
\cite{Kramer1}-\cite{Tsindlekht1}, in the framework of catastrophe
theory we investigate the equilibrium set for the system of strongly
correlated electron gas at the conditions of dHvA effect. We show
that in every period of dHvA oscillations the discontinuities of
order parameter accompanied DPT is handled by Riemann-Hugoniot
catastrophe implying the standard scenario for the transition, e.g.
DPT is of the second order at the center of dHvA period, weakly
first order in the nearest vicinity of this point and is of the
first order at the rest part of the dHvA period both on $\it
{temperature}$ and $\it {magnetic}$ $\it {field}$. Thus, similar to
other magnetic systems, e. g. the spin \cite{Hubert} and
metamagnetic ones \cite{Binz}, DPT can be realized in a number of
different nonuniform phases depending on the shape of the sample.

At $T=$ 0 K the fine, sub-band, structure of Landau level can
results in DPT of the first order on $\it {magnetic}$ $\it {field}$
\cite{Blanter}, but this effect based on treating the CD instability
as an electron topological transition is negligibly small in
comparing with the results obtained in the present publication.

We show that the condition for DPT occurrence $a=$1 where
$a=\mu_{0}\max\{\partial M/\partial B\}$ is a differential magnetic
susceptibility is valid only in the nearest vicinity of the center
of dHvA period, $x=$0, where $x$ is an increment of magnetic field
\cite{Shoenberg}. This condition is violated for $x\ne 0$ and is replaced by the generalized condition on $a$ in accordance with the bifurcation set of cusp
catastrophe. We discuss the conditions for realization of the DPT in
the sample with taking into account the long-range dipole-dipole
interaction and compare the results with available data.

\begin{figure} [t]
  \includegraphics[width=0.4\textwidth]{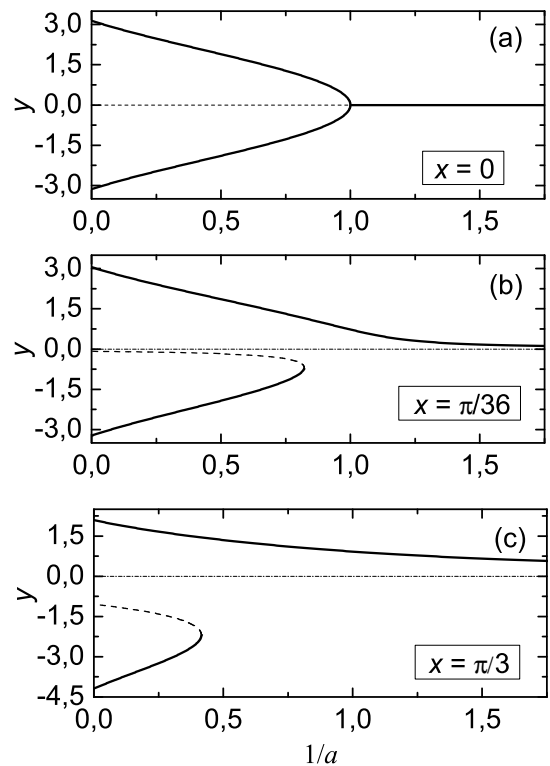}
\caption{\ Reduced magnetization $y$ as a function of inverse
differential magnetic susceptibility $1/a$ at three different values
of magnetic field $x=$0 (a), $\pi$/36 (b) and
$\pi$/3 (c). The solid (dash) lines correspond to the local minimum
(maximum) of the free energy density $G(y;a,x)$. Symmetric
supercritical pitchfork bifurcation (a) of the system undergoing the
second order phase transition at the center of dHvA period $x=0$ for
$a=1$ (critical point) is replaced by the imperfect bifurcation
diagram (b, c) with increase of $\mid x \mid \le \pi$ ($2 \pi$ is a
period of dHvA oscillations in reduced units) when the system is a
subject for the first order phase transition.} \label{Magnetization}
\end{figure}

\section{\label{sec:Model}Model}

Equilibrium properties of strongly correlated electron gas at the
conditions of dHvA effect in one-harmonic approximation can be
described by Gibbs free energy $G(y;a,x)=a \cos
{(x+y)}+\frac{1}{2}y^{2},$ which is a functional of two conjugated
variables $x$ and $y=-\partial_{x}G$ \cite{Shoenberg}. Here, the
small-scale field $x=k\mu_{0}(H-H_{a})\in [-\pi,+\pi]$ is the
increment of the large-scale internal magnetic field $\mu_{0}H$ and
applied magnetic field $\mu_{0}H_{a}$, $y$ is oscillating part of
reduced magnetization, $k=2\pi F/(\mu_{0}H_{a})^{2}=2\pi /\Delta H$,
$F$ is the fundamental frequency of the dHvA oscillations
corresponding to the extreme cross-section of Fermi-surface, and
$\Delta H$ is dHvA period. Minimization of the free energy
$G(y;a,x)$ with respect to $y$ leads to well-known expression for
the magnetization \cite{Shoenberg}
\begin{equation} \label{eq:magnetization}
y=a \sin {b} \qquad \\
\end{equation}
with the magnetic flux density $b=x+y$ being a function of increment
of internal magnetic field $x$ and reduced magnetization $y$.
Internal magnetic field $x=x_{0}+x_{ms}$ depends on applied magnetic
field $x_{0}$ and stray field $x_{ms}$, which take into account the
long-range dipole-dipole interaction dependent on the shape of a
sample and has to be found by solving the Maxwell equations in
magnetostatic approximation by a self-consistent procedure. As it
was suggested by Shoenberg \cite{Shoenberg}, the magnetic flux
dependence of oscillating part of magnetization $y$
Eq.~(\ref{eq:magnetization}) at curtain conditions, when the
differential susceptibility $a \ge $1, results in arising
instability of uniform phase (Shoenberg effect). At this condition
both the magnetization and magnetic induction become a multi-valued
function of internal magnetic field $x$. Therefore, depending on
long-range dipole-dipole interaction at high magnetic field and low
temperature the system of strongly correlated electron gas can
undergo the DPT with formation of IDP
which corresponds to a global minimum of the free energy of the
system. Due to the governing role of the long-range dipole-dipole
interactions, the morphology of the DPT depends crucially on the
shape of the sample which, in general, cannot be account for by the
straightforward use of conception of demagnetizing coefficient
\cite{Logoboy1}.

The magnetization $y$ Eq.~(\ref{eq:magnetization})
as a function of inverse reduced amplitude of the dHvA oscillations
$1/a$ at three values of the increment of magnetic field $x=$0,
$\pi/$3 and 2$\pi/$3 is plotted in Fig.~\ref{Magnetization}, which
illustrates the typical bifurcation behavior of the system, e. g.
the existence of two stable states for magnetization at some values
of the parameters $a \ge$1 and $x$, and the possibility of
discontinues change of the order parameter through the disappearance
of stable states when $x \ne 0$.

\section{\label{sec:Results and Discussions}Results and Discussions}

\subsection{\label{Condition of Diamagnetic Instability}Condition of Diamagnetic Instability}

\begin{figure}
  \includegraphics[width=0.4\textwidth]{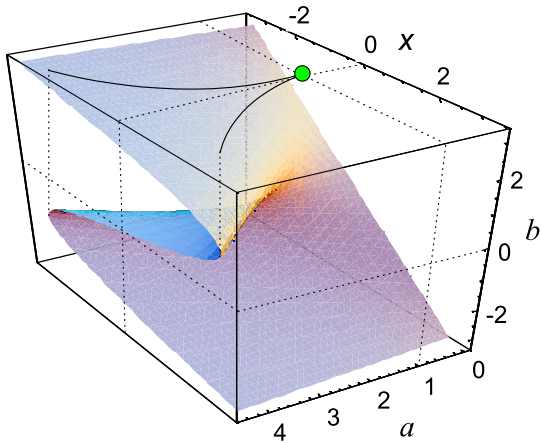}
\caption{\ (color online) Shown are the equilibrium surface and
bifurcation set (solid line) of cusp catastrophe for the system of
electron gas under conditions of strong dHvA effect. At crossing
the bifurcation set with change of control variables $a$ and $x$,
the system of strongly correlated electron gas is a subject for the
DPT of the first order. The only point $(a=1, x=0)$ where the system
undergoes the second-order phase transition, e. g. the critical
point, is a triple degenerate point shown by solid circle.}
\label{Equilibrium Manifold}
\end{figure}

The discontinuities of the function $b=b(a,x)$ where parameter $a$
controls the amount of ordering, or the value of order parameter,
and parameter $x$ breaks the $Z_{2}$ symmetry of order parameter,
can be handled in the framework of catastrophe theory. The standard
approach allows to obtain the canonical expression for the normal
form $f(\eta;\boldsymbol {\beta})=\beta_{1}-\beta_{2}\eta+\eta^{3}$
justifying the existence of cusp catastrophe in the problem
(Fig.~\ref{Equilibrium Manifold}). Here, the new control variables
$\beta_{1}$ and $\beta_{2}$ relate to the origin ones $(a,x)$ by
means of relations $\beta_{1}=-$6$^{4}xa^{-4}$ and
$\beta_{2}=$6$^{-3}(a-1)xa^{-4}$ correspondingly, and
$\eta=$6$a^{-1}b$ is the scaled phase-state variable. The direct
calculations lead to the explicit form of the bifurcation set

\begin{figure}[t]
  \includegraphics[width=0.4\textwidth]{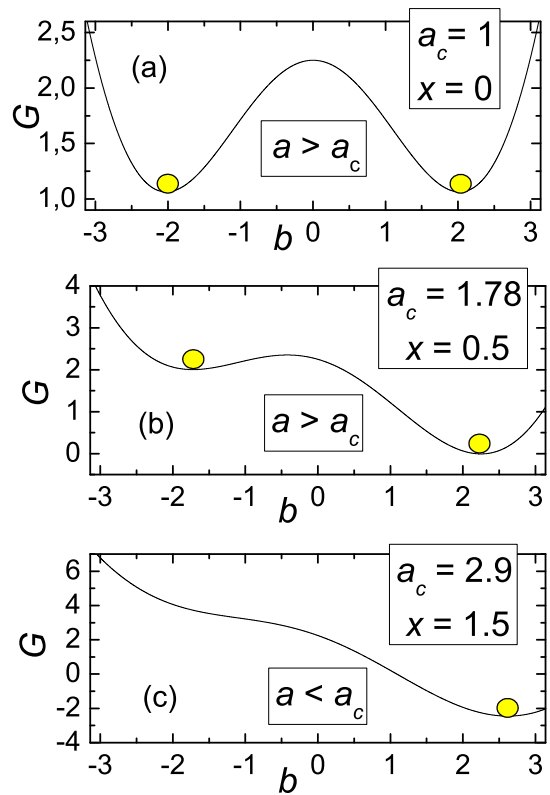}
\caption{ (color online) Free energy density $G=G(b)$ as a function of
magnetic induction for fix value $a=$2.25 and three different value
of $x=$0, 0.5 and 1.5 is shown for one dHvA period. (a) For $x=$0 and
$a_{c}=$1, the system has twofold degenerate ground state: two
minimums (circles) corresponding to the same value of $G$ can give
rise to the simple plane-parallel domain structure. (b,c) The increase of
$x$ followed by increase of $a_{c}$ breaks the equivalency of
the steady states, results in increase of the difference between the
the energies of two equilibrium states and possibility of
realization of variety of domain structures. For $a=$2.25 and
$x=$1.5 the only minimum implies the absolute stability of uniform
diamagnetic phase. }\label{Potential Function}
\end{figure}

\begin{equation} \label{eq:Bifurcation Set 1}
a\cos(\sqrt{a^{2}-1}-\mid x \mid)=1.
\end{equation}
defining the bifurcation curve on $(a,x)$ plane of control variables
where the fold bifurcation occurs. Eq.~(\ref{eq:Bifurcation Set 1})
is a generalized condition for DPT occurrence. It can be simplified in the vicinity of triple degenerate point
(a=1,x=0) resulting in standard cusp bifurcation set
\begin{equation} \label{eq:Bifurcation Set 2}
 8(1-a)^{3}+9x^{2}=0~
\end{equation}
which is a semicubic parabola. It follows from
Eq.~(\ref{eq:Bifurcation Set 1}) that the usually accepted condition
for DPT occurrence $a=$1 \cite{Shoenberg} is justified only at the
center of the dHvA period, $x=0$, but violated for
$x\ne$0 and replaced by
\begin{equation} \label{eq:DPT}
 a=a_{c}~,
\end{equation}
where $a_{c}$ is defined by Eq.~(\ref{eq:Bifurcation Set 1}) is a field-dependent. In particular, for $x \to$ 0, we obtain
\begin{equation} \label{eq:Critical Value}
 a_{c}=1+\frac{3^{2/3}}{2}x^{2/3}
\end{equation}

The character of bifurcation may be seen from the surface plot
$b=b(a,x)$, or the surface of stationary solutions of $b=x+a\sin
b$. The results of calculations are represented in
Fig.~\ref{Equilibrium Manifold} which shows the equilibrium manifold
of the system of strongly correlated electron gas for one period of
dHvA oscillations in joint space of state and control variables
$(b,a,x)$ and the bifurcation set Eq.~(\ref{eq:Bifurcation Set 1})
in $(a,x)$-plane. At the center of dHvA period ($x=0$) and
additional condition $a \to 1$ the system exhibits the symmetric
pitchfork bifurcation and undergoes the phase transition of the
second order from homogeneous to inhomogeneous state at decrease of
temperature $T$ (increase of $a$) with formation of CD structure the
type of which is governed by the long-range dipole-dipole
interaction and depends on the shape of the sample. The critical
point $a=1, x=0$, e. g. triple degenerate point, is the only point
where the DPT is of the second order. But, at $x \ne 0$, a
discontinues jump of order parameter implies the possibility of
existence of the first order DPT when the projection of equilibrium
state on $(a,x)$-plane crosses the bifurcation set with change of
temperature $T$ or magnetic field $x$ (see, also
Fig.~\ref{Magnetization}). A simple graphical illustration of
above-mentioned effects is provided by corresponding potential
function, $G(b;a,x)$, drawn as a function of magnetic induction $b$
for $a=$ 2.25 and three different value of $x$ in
Fig.~\ref{Potential Function} which shows the possibility of
formation of CD structure due to existence of two minima around the
center of dHvA period (Fig.~\ref{Potential Function}a,b) and the
absence of the condition for realization of IDP
for greater values of $x$ when two minima coalesce and the
uniform ground state is not degenerate (Fig.~\ref{Potential
Function}c).

\begin{figure}
  \includegraphics[width=0.4\textwidth]{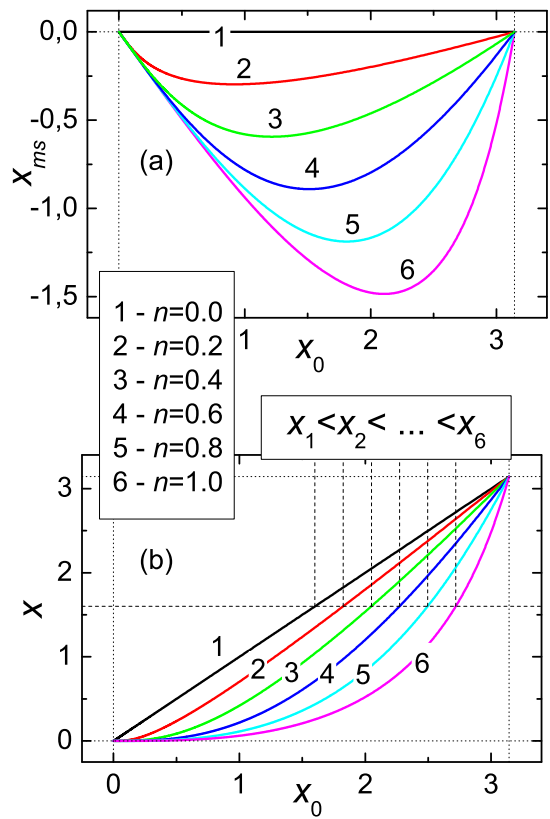}
\caption{ (color online)  (a) Magnetostatic field, $x_{ms}$, and (b)
internal magnetic fields, $x$, at crossing the bifurcation set from
high field side are plotted as a functions of applied field,
$x_{0}$, for different values of demagnetizing coefficient, $n$. At
given value of internal field, shown by horizontal dash line in (b),
the values $x_{i}$ ($i=$1-6) of applied field correspond to the
increasing values of $n$. Both function, $x_{ms}$ and $x$, are odd
functions of $x_{0}$, thus, only positive interval of $x_{0}$ is
shown. }\label{Internal vs Applied Fields}
\end{figure}

\subsection{\label{Shape Effect}Shape Effect}

The catastrophe theory allows to eliminate all the possible stable
equilibria (local minima) for a given pair of control variables
($a,x$) with the magnetic induction $b$, as a phase-state variable,
and establish the boundaries of CDs instability. Thus, the
straightforward application of the results consists in evaluation of
the range of existence of IDP, $\Delta x_{0}$, defined as a part of
dHvA period occupied by CDs. The boundary between different phases
is governed by the bifurcation set of the cusp catastrophe. Due to
the long-range character, the dipole-dipole interaction is sensitive
to the shape of a sample which results in the dependence of the
internal magnetic field $x$ on experiment arrangement. The simplest
way to account for the shape of the sample is the use of
demagnetizing coefficient $n$ for uniformly magnetized ellipsoid
subjected to the uniform magnetic field along principal axis. In
this case, the magnetization and magnetic induction are
single-valued functions of applied magnetic field, $x_{0}$, the
ground state of the system is uniform and the shape effects can be
calculated numerically. This procedure allows to evaluate the
conditions of occurrence of the diamagnetic instability. Of cause,
the morphology of the domain patterns in the IDP cannot be account
for by the use of the conception of demagnetizing coefficient, but
has to be established by solving Maxwell equations. The results of
calculation of the stray field, $x_{ms}$, as well as the internal
field, $x$, as a function of applied field, $x_{0}$, for several
values of $n$ when the uniform magnetization goes to the value
defined by bifurcation set Eq.~(\ref{eq:Bifurcation Set 1}) are
plotted in Fig.~\ref{Internal vs Applied Fields} which illustrates
the well-known effect, e. g. the increase of $n$ results in the
growing difference between internal and applied magnetic fields due
to the increase of absolute value of stray fields. In particular,
considering the samples with different demagnetizing coefficients,
it follows from Fig.~\ref{Internal vs Applied Fields}(b), that at
other equal conditions, the phase boundary is reached at the higher
values of applied field for the sample with the larger value of $n$.
Therefore, IDP occupies the larger part of the dHvA period for the
sample with $n \to 1$ which are preferable for observation of CD
instability. In connection to this, it is worth to be mentioned that
the discovery of the Condon domains was made on a plate-like sample
of Ag \cite{Condon_Walstedt} and the diamagnetic instability was
observed mostly by use the sample with the values of relevant
demagnetizing coefficient in the range $n \approx$ 0.5 - 1 (see,
\cite{Kramer1}, \cite{Solt3}).

The effect of "broadening" of IDP when the phase boundary is drawn
in ($\alpha_{1},x_{0}$) plane is illustrated by Fig.~\ref{Phase
Boundary} which shows the first-order phase curves for several
values of demagnetizing coefficient. In the calculation we neglected
the overheating-undercooling effects.

\begin{figure}[b]
  \includegraphics[width=0.4\textwidth]{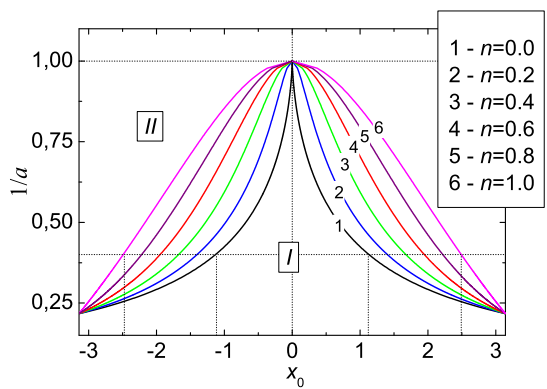}
\caption{\ (color online) The phase boundary is shown for different
values of demagnetizing coefficient $n$ in one period of dHvA
oscillations. At given value of $n$, the inner (outer) region,
designated by $I$ ($II$), corresponds to non-uniform (uniform)
diamagnetic phase. For fix $a$ (horizontal dash line) the
part of dHvA period occupied by IDP is wider for the sample with
larger demagnetizing coefficient.} \label{Phase Boundary}
\end{figure}

\begin{figure}[t]
  \includegraphics[width=0.4\textwidth]{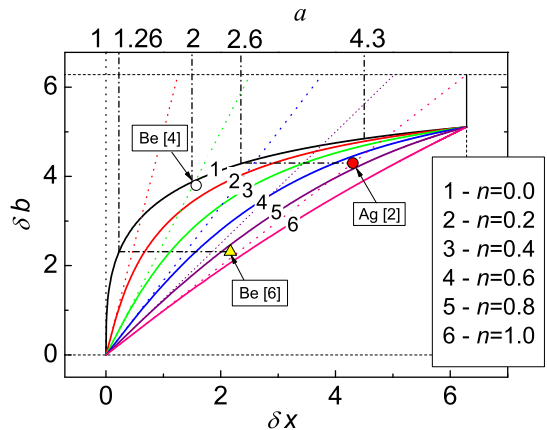}
\caption{ (color online) Family of curves $\delta b=\delta b(\Delta
x,n)$ is shown for one dHvA period. The slope of the curves at
$\delta x \to 0$ is defined by inverse demagnetizing coefficient,
1$/n$, in accordance with Eq.~(\ref{eq:Slope}) justifying the
Shoenberg theory of shape effects at the vicinity of critical point
($a$=1, $x$=0). The symbols accompanied by reference number
represent the experimental data.} \label{IDP Parameters}
\end{figure}

%\subsection{\label{Data Analysis}Data Analysis}
To compare the results of the theory with available experimental
data on observation of CDs instability we calculate the
characteristics of IDP. The formation of IDP is characterized by two
measured principal parameters, the average magnetic induction
splitting, $\delta b$, defined as the difference, calculated at the
center of dHvA period, between the values of magnetic induction in
two adjacent domains and the range of existence of CDs, $\delta x$,
defined as a part of dHvA period occupied by non-uniform phase
\cite{Logoboy2}. Though the occurrence of IDP and the existence of
CDs itself depends on long-range dipole-dipole interaction, the
average value of magnetic induction splitting, $\delta b$, depends
on the differential magnetic susceptibility and with reasonable
accuracy is defined in a explicit form by \cite{Shoenberg}
\begin{equation} \label{eq:Splitting}
\delta b-2a\sin \frac{\delta b}{2}=0.
\end{equation}
As far as concerned the range of existence of IDP $\delta x$, it is
a shape-dependent quality
due to dependence of internal field on dipole-dipole
interaction sensitive to the sample shape, as illustrated in
Fig.~(\ref{Internal vs Applied Fields}). Inasmuch as both
characteristics of IDP originate from the same
Eq.~(\ref{eq:magnetization}), they can be related to each other by
eliminating  $a$. It
results in rather complicated functional dependence which is handled
numerically, but in a limit $\delta x, \delta b \to$ 0 has a simple
form of linear function
\begin{equation} \label{eq:Slope}
 \delta b \approx \frac{1}{n} \delta x. \qquad \qquad
\end{equation}
From mathematical point of view Eq.~(\ref{eq:Slope}) is fulfilled in
an infinitely small vicinity of the triple degenerate point ($x=$0,
$a=$ 1), justifying the applicability of Shoenberg theory of shape
effect at this critical point. Accounting for the shape of the
sample we calculate the family of the curves $\delta b=\delta
b(\delta x,n)$ plotted in Fig.~\ref{IDP Parameters}. The graphical
representation of function $\delta b=\delta b(\delta x,n)$ which
relates the characteristics of IDP to the shape of the sample is
convenient to use for comparing the theoretical results with
experiment. For this, we introduce also the characteristics of IDP
which can be measured (or evaluated) directly in experiment, $\delta
H=\delta x/k$ and $\delta B=\delta x/k$.

The formation of IDP is accompanied by irreversible behavior of the
magnetization curve which can be used for experimental investigation
of IDP. The experimental detection of diamagnetic hysteresis was
performed on Be single crystal rod-like sample of size
8$\times$2$\times$1 mm$^{3}$ \cite{Kramer2}. The  prolate ellipsoid
which approximates the sample has demagnetizing coefficient $n=$
0.04. The hysteresis was observed at $\approx$0.25 part of the dHvA
period which is 2$\pi$ in reduced units. It is reasonable to
associate the range of existence of IDP with the range of existence
of hysteresis. Thus, we obtain $\delta x \approx$1.57.
Unfortunately, the magnetic induction splitting is not reported in
experiment, but it can be calculated by use of the value $a$ due to
Eq.~(\ref{eq:Splitting}). At the conditions of experiment, $T=$1.3
K, $T_{D}=$2 K (Dingle temperature) and $\mu_{0}H_{a}=$3.6 T, the
differential magnetic susceptibility calculated  in the model of
slightly corrugated Fermi surface \cite{Logoboy3} is $a \approx$2
which with the aid of Eq.~(\ref{eq:Splitting}) gives $\delta b
\approx$3.8. To verify the consistency of this value of $\delta b$
we calculate the absolute value of induction splitting $\delta B
\approx$78.6 G using the reported value of dHvA period $\Delta H
=$130 G and compare it with the experimentally detected global
change of induction $\Delta B\approx$121 G at one dHvA period. The
inequality $\delta B < \Delta B$ confirms the reliability of
calculated value of $\delta b \approx$3.8. The point corresponding
to the values of $\delta x \approx$1.57 and $\delta b \approx$3.8 is
plotted in Fig.~\ref{IDP Parameters} and lies at the curve for $n
\approx$0.04, as it is expected.

The series of experiments on observation of CDs by $\mu$SR
spectroscopy \cite{Solt1}-\cite{Solt3} have proved the occurrence of
diamagnetic instability of strongly correlated electron gas in a
number of normal metals, revealed the possibility of the first-order
phase transition by detecting the overheating-undercooling effects
and actually served as one of the stimulation factors for
development of the theory. As an example, we consider the data for
Be \cite{Solt1} which provide the value of magnetic induction
splitting and range of existence of CDs with high accuracy of
$\approx \pm$1 G in applied field of several Tesla. The $\mu$SR
measurements in \cite{Solt1} were performed on plate-like samples of
size 1$\times$1$\times$8 mm$^{3}$. At the conditions of experiment,
$T=$0.8 K and $\mu_{0}H_{a}=$2.75 T, the experimental value of
induction splitting $\delta B =$ 28.8$\pm$1.4 G which gives $\delta
b=$ 2.31 with use of the reported dHvA period $\Delta H=$78.2 G. The
value of the range of existence of IDP is $\delta H \approx$27 G or
$\delta x \approx$2.2. The plotted in Fig.~\ref{IDP Parameters}
point, $\delta x \approx$2.2 and $\delta b \approx$2.31, belongs to
the curve with $n \approx$0.8 which is close to the theoretical
value of $n=$0.75 for oblate ellipsoid approximating the sample.

The famous experiment on observation of CDs instability by NMR
technique \cite{Condon_Walstedt} on the plate-like sample of Ag of
size 8$\times$8$\times$0.8 mm$^{3}$ at helium temperature and
applied field $\mu_{0}H_{a}=$9 T provides both values $\delta
B=$$\delta H\approx$ 11 which allows to calculate the reduced values
$\delta b=$$\delta x\approx$ 4.32 with use of experimental value of
dHvA period $\Delta H=$15.9 G. The corresponding point plotted in
Fig.~\ref{IDP Parameters} lies at the curve parameterized by $n
\approx$ 0.78 which is close to the value $n \approx$ 0.84 obtained
for inscribed oblate ellipsoid approximated the sample. Thus, there
is a good agreement of the theoretical results with the experimental
data.

The absence of data at the vicinity of origin (Fig.~\ref{IDP
Parameters}) definitely calls for further experimental investigation
of diamagnetic instability of strongly correlated electron systems.

One of the properties of the family of curves $\delta b=\delta
b(\delta x,n)$ Fig.~\ref{IDP Parameters} consists in its
universality in a sense that the curves being a graphical
representation of the relationship between the measured properties
of IDP (order parameter $\delta b$ and range of existence of IDP
$\delta x$) and characteristics of the shape of the sample
(demagnetizing coefficient $n$) do not depend on concrete Fermi
surface. Therefore, the data related to different systems exhibiting
the diamagnetic instability can belong to the same curve. One can
conclude that plotted in the ($\delta b, \Delta$) plane where
$\Delta$ is the range of existence of IDP in terms of internal
magnetic field, the family of curves (Fig.~\ref{IDP Parameters})
gathers onto a single universal, or parent curve $\delta b=\delta
b(\delta x,n=0)$. This is a consequence of the universality of the
bifurcation theory in description of diamagnetic instability of the
electron system in a joint space (b,a,x) of phase and control
variables. Another feature of the dependencies represented in
Fig.~\ref{IDP Parameters} consists in a possibility of calculation
the value of $a$ for given experimental arrangement. As it follows
from Eq.~(\ref{eq:Splitting}), a straight horizontal line in
Fig.~\ref{IDP Parameters} crosses the family of curves in points of
equal $a$. Mapping the points belonging to different curves onto the
parent curve $\delta b=\delta b( \Delta)$ parameterized by $a$, with
careful tabulation provides the values of $a$ which can be compared
with corresponding values calculated in different models of Fermi
surface. Some examples of the mapping of the experimental points
onto parent curve are shown in Fig.~\ref{IDP Parameters} by dash dot
lines. It is follows from the analysis of the function $\delta
b=\delta b(\delta x,n)$ that predicted dependency on the shape of
the sample can be detectable for moderate values of reduced
amplitude of dHvA oscillations $a \approx$ 1.3-3. For greater values
of $a$ the high precision experiments are needed due to the small
difference between the curves.

\section{\label{sec:Conclusions}Conclusions}

The diamagnetic instability and stable equilibria of the system of
the strongly correlated electron gas at high magnetic field and low
temperature are studied by the tools of catastrophe theory. While at
the center of dHvA period the symmetric supercritical pitchfork
bifurcation gives rise to the second order phase transition on
temperature, the deviation of magnetic field from the value
corresponding to the center of dHvA period results in the change of
the phase transition order from the second to the weakly first in
the nearest vicinity of triple degenerate point and first at the
rest part of dHvA period both on temperature and magnetic field. We
calculate the generalized condition for DPT occurrence.

The results of our investigation show a reasonable agreement with
the experiments. We believe also that the observation of DPT and
interpretation of data can be complicated by the intrinsic structure
of IDP which can realized through band patterns, bubbles and so on,
but this is beyond the scope of the present publication.

We hope that our studies provide consistent framework
in consideration of diamagnetic instability of strongly correlated
electron gas, will help in making a quantitative systematic analysis
of experimental data and stimulate the further experiments on
investigation of the diamagnetic instability.

\begin{acknowledgments}
We are indebted to V. Egorov, R. Kramer and I. Sheikin for fruitful
discussions.
\end{acknowledgments}

\end{document}